\documentclass[fleqn,twoside,psfig]{article}
\usepackage{espcrc2}

\usepackage{psfig}
\newcommand{\bei}{\begin{itemize}}
\newcommand{\eei}{\end{itemize}}
\newcommand{\beq}{\begin{equation}}
\newcommand{\eeq}{\end{equation}}
\newcommand{\beqn}{\begin{eqnarray}}
\newcommand{\eeqn}{\end{eqnarray}}
\newcommand{\beqns}{\begin{eqnarray*}}
\newcommand{\eeqns}{\end{eqnarray*}}
\newcommand{\vs}{\\[0.3cm]\indent}

\newcommand{\intl}{\int\limits}

\newcommand{\mc}{\multicolumn}

\RequirePackage{xspace}
\usepackage{relsize}
\def\babar{\mbox{\slshape B\kern-0.1em{\smaller A}\kern-0.1em
    B\kern-0.1em{\smaller A\kern-0.2em R}}}
\def\pc{$\%$}

\def\sf{spectral function}
\def\sfs{spectral functions}

\def\ee{$e^+e^-$}

\def\ee{$e^+e^-$}

\def\amuhadLO{$a_\mu^{\rm had,LO}$}
\def\rar{\rightarrow}
\def\via{via} 

\def\cf{{\em cf.}}
\def\rs{\raisebox{1.5ex}[-1.5ex]}

\title{Updated Estimate of the Muon Magnetic Moment \\[0.1cm]
Using Revised Results from \boldmath\ee\ Annihilation}

\author{Michel Davier\\
	Laboratoire de l'Acc\'el\'erateur Lin\'eaire\\
        IN2P3/CNRS et Universit\'e de Paris-Sud\\
        91898 Orsay, France\\
	E-mail: davier@lal.in2p3.fr}

\begin{document}

\begin{abstract}
A new evaluation of the hadronic vacuum polarization contribution 
to the muon magnetic moment is presented. We take into account
the reanalysis of the low-energy \ee\ 
annihilation cross section into hadrons by the CMD-2 Collaboration.
The agreement between \ee\ and $\tau$ spectral functions in the
$\pi\pi$ channel is found to be much improved. Nevertheless, significant 
discrepancies remain in the center-of-mass energy range between 0.85 and 
$1.0\;{\rm GeV}/c^2$.The deviations from the  
measurement at BNL are found to be $(22.1 \pm 7.2 \pm 3.5 \pm 8.0)~10^{-10}$ 
(1.9~$\sigma$) and $(7.4 \pm 5.8 \pm 3.5 \pm 8.0)~10^{-10}$ (0.7~$\sigma$) 
for the \ee- and $\tau$-based estimates, respectively, where the second 
error is from the nonhadronic contributions and the third one from 
the BNL measurement. Taking into account the $\rho^- - \rho^0$ mass
splitting determined from the measured spectral functions increases
the $\tau$-based estimate and leads to a worse discrepancy between 
the two estimates.
\vspace{1pc}
\end{abstract}

\maketitle
\setcounter{page}{1}
%
% ----------------------------- Introduction ----------------------------
%
\section{Introduction}
\label{sec_introduction}

Hadronic vacuum polarization in the photon propagator plays an important 
role in the precision tests of the Standard Model. This is the case for 
the muon anomalous magnetic moment $a_\mu=(g_\mu -2)/2$ where the
hadronic vacuum polarization component, computed from experimentally
determined spectral functions, is the leading contributor to the
uncertainty of the theoretical prediction.
\vs
Spectral functions are directly obtained from the  cross sections for \ee\
annihilation into hadrons. The accuracy of the calculations has therefore
followed the progress in the quality of the corresponding data~\cite{eidelman}.
Because the latter was not always suitable, it was deemed necessary to resort 
to other sources of information. One such possibility was the 
use of the vector spectral functions~\cite{adh} derived from the study 
of hadronic $\tau$ decays~\cite{aleph_vsf} for the energy range less 
than $m_\tau\sim1.8~{\rm GeV}$. 
Also, it was demonstrated that perturbative QCD could be 
applied to energy scales as low as 1-2~GeV~\cite{aleph_asf},
thus offering a way to replace poor \ee\ data in some energy regions 
by a reliable and precise theoretical 
prescription~\cite{dh97,steinhauser,martin,groote,dh98}. 
\vs
A complete analysis including all available experimental data was 
presented in Ref.~\cite{dehz}, taking advantage of the new precise 
results in the
$\pi\pi$ channel from the CMD-2 experiment~\cite{cmd2} and from the
ALEPH analysis of $\tau$ decays~\cite{aleph_new}, and benefiting 
from a more complete
treatment of isospin-breaking corrections~\cite{ecker1,ecker2}. 
In addition to these major updates, the contributions of the many exclusive
channels up to 2~GeV center-of-mass energy were completely revisited. It was
found that the \ee\ and the isospin-breaking corrected $\tau$ spectral 
functions were not consistent within their respective uncertainties, 
thus leading to inconsistent predictions for the  lowest-order 
hadronic contribution to the muon magnetic anomaly. The leading contribution 
to the discrepancy originated in the $\pi\pi$ channel
with a difference of $(-21.2\pm6.4_{\rm exp}\pm2.4_{\rm rad}\pm2.6_{\rm SU(2)}
\,(\pm7.3_{\rm total}))~10^{-10}$. The estimate based on \ee\ data agreed 
with another analysis using the same input data~\cite{teubner}. 
When compared to the world average of the muon magnetic anomaly
measurements,
\beq
\label{eq:bnlexp}
	a_\mu \:=\: (11\,659\,203 \pm 8)~10^{-10}~,
\eeq
which is dominated by the 2002 BNL result using positive 
muons~\cite{bnl_2002},
the respective \ee-based and $\tau$-based predictions disagreed at the 3.0 
and 0.9 $\sigma$ level, respectively, when adding experimental and 
theoretical errors in quadrature.
\vs
Our analysis had to be updated~\cite{dehz03} since
the CMD-2 Collaboration at Novosibirsk discovered that part of the
radiative treatment was incorrectly applied to the data and produced a 
complete reanalysis~\cite{cmd2_new}. As the CMD-2 data 
dominate the \ee-based prediction, the changes produce a significant 
effect in the final result.
No significant change occurred for the $\tau$-based prediction. The
only relevant fact is a new result~\cite{L3_hpi0} for the branching 
ratio of the $\tau^- \rightarrow \nu_\tau h^- \pi^0$ mode 
($h^-$ stands for a charged pion or kaon). 

%
% ----------------  Muon Magnetic Anomaly -------------------
%
%
% --------- 
%
\section{Muon Magnetic Anomaly}
\label{anomaly}

It is convenient to separate the Standard Model (SM) prediction for the
anomalous magnetic moment of the muon
into its different contributions,
\beq
    a_\mu^{\rm SM} \:=\: a_\mu^{\rm QED} + a_\mu^{\rm had} +
                             a_\mu^{\rm weak}~,
\eeq
with
\beq
 a_\mu^{\rm had} \:=\: a_\mu^{\rm had,LO} + a_\mu^{\rm had,HO}
           + a_\mu^{\rm had,LBL}~,
\eeq
and where $a_\mu^{\rm QED}=(11\,658\,470.6\pm0.3)~10^{-10}$ is 
the pure electromagnetic contribution (see~\cite{hughes,cm} and references 
therein~\footnote
{
	Some adjustment was recently made concerning the fourth-order 
	contribution from the leptonic light-by-light scattering, mostly 
	affecting the QED prediction for $a_e$ and through it the value 
	of $\alpha$~\cite{kino_nio,nyff}. The resulting change in 
	$a_\mu^{\rm QED}$ is within the quoted uncertainty of $0.3~10^{-10}$ 
	and has not been included in the present analysis.
}), \amuhadLO\ is the lowest-order contribution from hadronic 
vacuum polarization, $a_\mu^{\rm had,HO}=(-10.0\pm0.6)~10^{-10}$ 
is the corresponding higher-order part~\cite{krause2,adh}, 
and $a_\mu^{\rm weak}=(15.4\pm0.1\pm0.2)~10^{-10}$,
where the first error is the hadronic uncertainty and the second
is due to the Higgs mass range, accounts for corrections due to
exchange of the weakly interacting bosons up to two loops~\cite{amuweak}. 
For the light-by-light (LBL) scattering part 
we add the values for the pion-pole 
contribution~\cite{knecht_light,kino_light_cor,bij_light_cor} and the
other terms~\cite{kino_light_cor,bij_light_cor} to obtain
$a_\mu^{\rm had,LBL}=(8.6\pm3.5)~10^{-10}$.
\vs
Owing to the analyticity of the 
vacuum polarization correlator, the contribution of the hadronic 
vacuum polarization to $a_\mu$ can be calculated \via\ the dispersion 
integral~\cite{rafael}
\beq\label{eq_int_amu}
    a_\mu^{\rm had,LO} \:=\: 
           \frac{\alpha^2(0)}{3\pi^2}
           \intl_{4m_\pi^2}^\infty\!\!ds\,\frac{K(s)}{s}R(s)~,
\eeq
where $K(s)$ is a well-known QED kernel.
In Eq.~(\ref{eq_int_amu}), $R(s)\equiv R^{(0)}(s)$ 
denotes the ratio of the 'bare' cross
section for \ee\ annihilation into hadrons to the pointlike muon-pair cross
section. The 'bare' cross section is defined as the measured cross section,
corrected for initial-state radiation, electron-vertex loop contributions
and vacuum polarization effects in the photon propagator (note that
photon radiation in the final state (FSR) is included in the 'bare' 
cross section). 
The reason for using the 'bare' ({\it i.e.} lowest order) 
cross section is that a full treatment of higher orders is anyhow 
needed at the level of $a_\mu$, so that the use of 'dressed' 
cross sections would entail the risk of double-counting some of the 
higher-order contributions.
\vs
The function $K(s)$ decreases monotonically with increasing $s$. It gives
a strong weight to the low energy part of the integral~(\ref{eq_int_amu}).
About 91\pc\ of the total contribution to \amuhadLO\ 
is accumulated at center-of-mass 
energies $\sqrt{s}$ below 1.8~GeV and 73\pc\ of \amuhadLO\ is covered by 
the two-pion final state which is dominated by the $\rho(770)$ 
resonance. 

%
% ------------------- The Data ---------------------------
%

\section{Changes to the Input Data}
\label{sec_data}

%
% ------------------------- The ee Data at low energies -----------------
%
%\subsection{\it \ee\ Annihilation Data}
%\label{sec_dat_ee}

\begin{figure}[t]
\centerline{\psfig{file=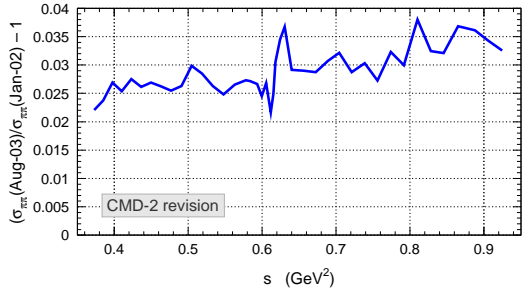,width=70mm}}
\caption[.]{\it 
	Relative change in the $e^+e^-\rightarrow\pi^+\pi^-$ cross section 
	of the revised CMD-2 analysis~\cite{cmd2_new} with respect 
	to the one previously published~\cite{cmd2}. }
\label{fig:cmd2_comp}
\end{figure}
The CMD-2 data, published in 2002 for the $\pi\pi$ channel~\cite{cmd2},
have been completely reanalyzed~\cite{cmd2_new} following the discovery 
of an incorrect implementation of radiative corrections in the analysis 
program. Overall, the pion-pair cross section increased by
$2.1\%$ to $3.8\%$ in the measured energy range 
(\cf\  Fig.~\ref{fig:cmd2_comp}), well above the previously
quoted total systematic uncertainty of $0.6\%$. Specifically, 
the leptonic vacuum polarization contribution in the t-channel had been 
inadvertently left out in the calculation of the Bhabha cross section. 
This effect produced a bias in the luminosity determination, 
varying from $2.2\%$ to $2.7\%$ in the 0.60-0.95~GeV energy range.
The problem consequently affected the measured cross sections for all
hadronic channels. Another problem was found in the radiative corrections
for the muon-pair process, ranging from $1.2\%$ to $1.4\%$ in the same region. 
A more refined treatment of hadronic vacuum polarization was performed, 
with changes not exceeding $0.2\%$ for most data points.
The effects in the Bhabha- and muon-pair channels also
affected the event separation and the measured ratio of pion pairs to 
electron and muon pairs changed by typically $0.7\%$. 
\vs
The correction of the bias in the luminosity determination increases all
hadronic cross sections published by CMD-2. The changes are $2.4\%$ and
$2.7\%$ on the $\omega$ and $\phi$ resonance cross sections, respectively. 
\vs
New published data by SND on the $\omega$ resonance~\cite{snd_omega} 
and the $2\pi^+2\pi^-$ as well as $\pi^+\pi^-2\pi^0$ 
modes~\cite{snd_4pi} (unchanged cross sections for the latter two, 
but reduced systematics with respect 
to previous publications) have been included in this update.  
\vs
A detailed discussion of radiative corrections, in particular the effect 
of final-state radiation by the charged hadrons was given in Ref.~\cite{dehz}.
Also given therein is a compilation of all input data (with references) 
used to calculate the integral~(\ref{eq_int_amu}).

%
% ------------------------------ CVC tests ---------------------
%
\section{Comparison of \ee\ and $\tau$ Spectral Functions}

\begin{figure}[p]
\centerline{\psfig{file=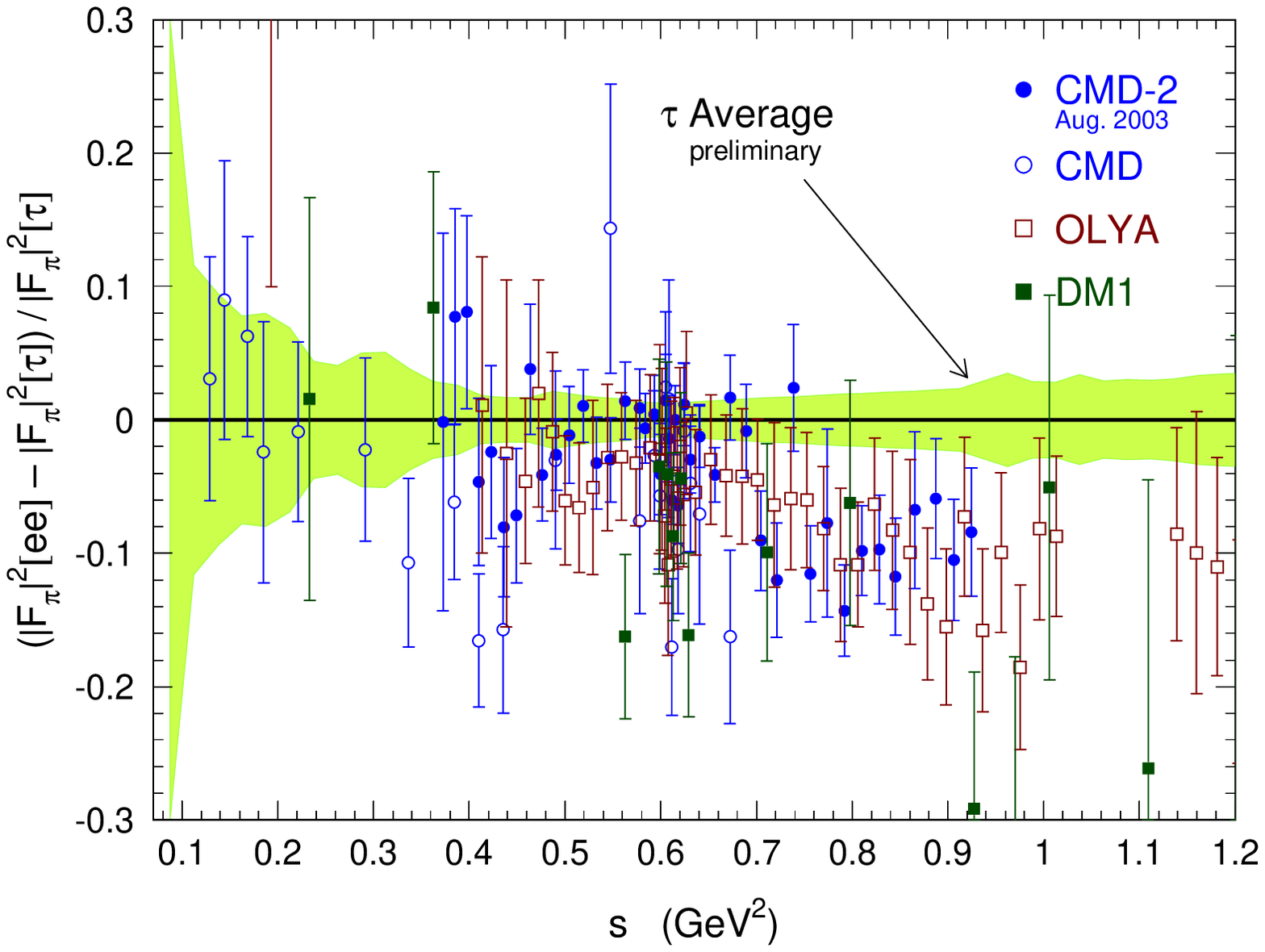,width=70mm}}
\caption[.]{\it Relative comparison of the $\pi^+\pi^-$ \sfs\
    	from \ee\  and isospin-breaking corrected $\tau$ data, 
	expressed as a ratio to the $\tau$ \sf.
	The band shows the uncertainty on the latter.
        The \ee\ data are from CMD-2~\cite{cmd2_new},
        CMD, OLYA and DM1 (references quoted in \cite{dehz03}).}
\label{fig_2pi_comp}
%\end{figure}
%\begin{figure}[t]
\vspace{0.6cm}
\centerline{\psfig{file=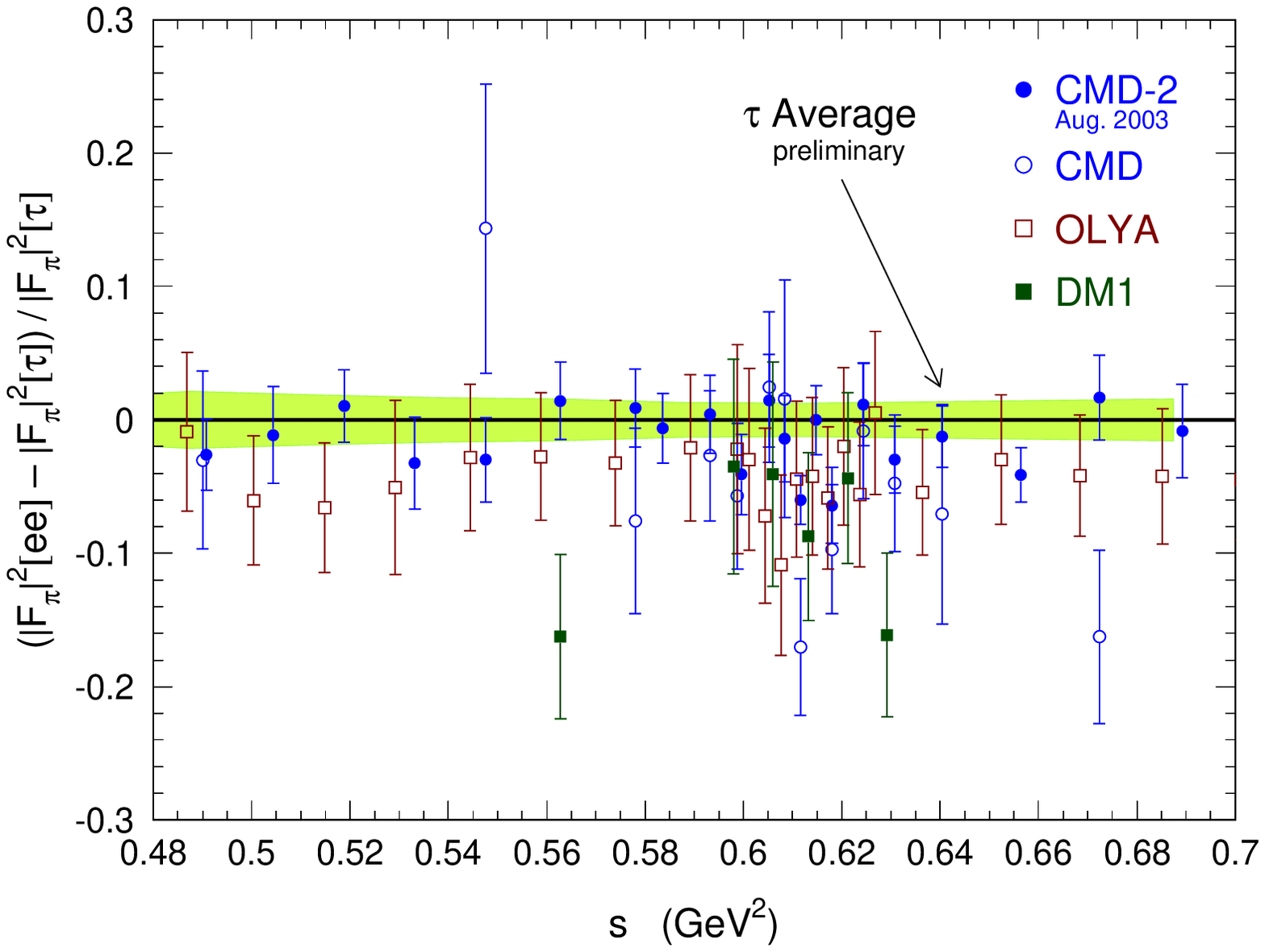,width=70mm}}
\caption[.]{\it Relative comparison in the $\rho$ region of the 
	$\pi^+\pi^-$ \sfs\
    	from \ee\  and isospin-breaking corrected $\tau$ data, 
	expressed as a ratio to the
     	$\tau$ \sf. The band shows the uncertainty on the latter.
        The references for \ee\ data are given in Fig.~\ref{fig_2pi_comp}.}
\label{fig_2pi_comp_zoom}
\end{figure}
The new \ee\ and the isospin-breaking corrected $\tau$ \sfs\ can be 
directly compared for the $\pi\pi$ final state. The $\tau$ \sf\ is obtained 
by averaging ALEPH~\cite{aleph_vsf}, CLEO~\cite{cleo_2pi} and
OPAL~\cite{opal_2pi} results~\cite{dehz}. The \ee\ data are plotted
as a point-by-point ratio to the $\tau$ \sf\ in Fig.~\ref{fig_2pi_comp}, 
and enlarged in Fig.~\ref{fig_2pi_comp_zoom}, to better emphasize the 
region of the $\rho$ peak. The central bands in Figs.~\ref{fig_2pi_comp} 
and \ref{fig_2pi_comp_zoom} give the quadratic sum of the statistical 
and systematic errors of the $\tau$ spectral function obtained 
by combining all $\tau$ data. 
The \ee\ data have moved closer to the $\tau$ results: 
they are now consistent below and around the peak, while, albeit 
reduced, the discrepancy persists for energies larger than 0.85~GeV.
\vs
A convenient way to assess the compatibility between \ee\ and $\tau$
\sfs\ proceeds with the evaluation of $\tau$ decay fractions using
the relevant \ee\ \sfs\ as input. All the isospin-breaking corrections 
detailed in Ref.~\cite{dehz} are included. 
This procedure provides a quantitative comparison 
using a single number. The weighting of the \sf\ is however different 
from the vacuum polarization kernels. Using the branching fraction
$B(\tau^-\rightarrow \nu_\tau\,e^-\,\bar{\nu}_e)\,=\,(17.810 \pm 0.039)\%$,
obtained assuming leptonic universality in the charged weak 
current~\cite{aleph_new}, the result for the $\pi\pi$ channel is
\beq
\label{eq:Bcvc}
{\cal B}_{\rm CVC}^{\pi \pi^0} = \\
	(24.52 \pm 0.26_{\rm exp} 
	  \pm 0.11_{\rm rad} \pm 0.12_{\rm SU(2)})\%~,
\eeq
where the errors quoted are split
into uncertainties from the experimental input
(the \ee\ annihilation cross sections) and the numerical integration procedure,
the missing radiative corrections applied to the relevant \ee\ data,
and the isospin-breaking corrections when relating $\tau$ and \ee\
\sfs. 
\begin{figure}[t]
\centerline{\psfig{file=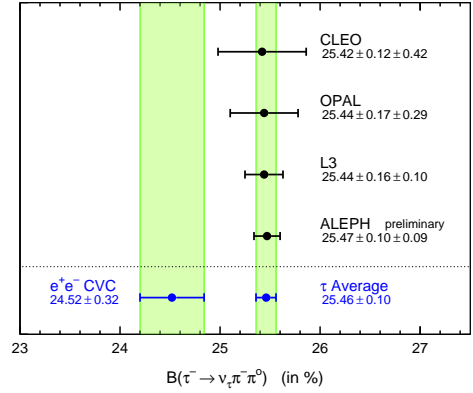,width=70mm}}
\caption[.]{\it The measured branching ratios for 
       $\tau^-\rightarrow\nu_\tau\pi^-\pi^0$ compared to the prediction
       from the $e^+e^-\rar\pi^+\pi^-$ \sf\ applying the isospin-breaking
       correction factors discussed in Ref.~\cite{dehz}.
       The measured branching ratios are from ALEPH~\cite{aleph_new},
       CLEO~\cite{cleo_bpipi0} and OPAL~\cite{opal_bpipi0}.
       The L3 and OPAL results are obtained from their $h \pi^0$ branching 
       ratio, reduced by the small $K \pi^0$ contribution 
       measured by ALEPH~\cite{aleph_ksum} and CLEO~\cite{cleo_kpi0}.}
\label{fig_cvc_2pi}
\end{figure}
Even though the corrections to the CMD-2 results have reduced 
the discrepancy between~(\ref{eq:Bcvc}) and the world average of 
the direct ${\cal B}(\tau^-\rightarrow \nu_\tau\,\pi^-\pi^0)$ 
measurements from 4.6 to 2.9 standard deviations (adding 
all errors in quadrature), the remaining difference of 
$(-0.94\pm0.10_\tau\pm0.26_{\rm ee}\pm0.11_{\rm rad}
\pm0.12_{\rm SU(2)}(\pm0.32_{\rm total}))\%$ is still problematic.
% where the uncertainties are from the $\tau$ branching ratio, 
% \ee\ cross sections, \ee\ missing radiative corrections and isospin-breaking 
% corrections (including the uncertainty on $V_{ud}$), respectively. 
Since the disagreement between \ee\ and $\tau$ \sfs\ is 
more pronounced at energies above 850~MeV, we expect a smaller discrepancy
in the calculation of \amuhadLO\ because of the steeply falling function
$K(s)$. More information on the comparison is
displayed in Fig.~\ref{fig_cvc_2pi} where it is clear that ALEPH, CLEO, L3 
and OPAL all separately, but with different significance,  
disagree with the \ee-based CVC result.

%
% ----------------------------- Results ------------------------------
%
\section{Results}
\label{sec_results}

%
% ------------
%
%\subsection{\it Lowest Order Hadronic Contributions}

The integration procedure and the specific contributions -- near $\pi\pi$
threshold, the $\omega$ and $\phi$ resonances, the narrow quarkonia and
the high energy QCD prediction -- are treated as in our previous
analysis~\cite{dehz}.
\begin{table*}[p]
\begin{center}
\setlength{\tabcolsep}{0.2pc}
{\small
\begin{tabular}{lcrrr} \hline 
&&&& \\[-0.2cm]
 & & \mc{3}{c}{\amuhadLO\ ($10^{-10}$)} \\
\rs{Modes} & \mc{1}{c}{\rs{Energy [GeV]}} & \mc{1}{c}{~\ee} 
	& \mc{1}{c}{$~\tau$\,$^(\footnotemark[3]{^)}$} 
		& \mc{1}{c}{$~\Delta(e^+e^--\tau)$} \\[0.15cm]
\hline
&&&& \\[-0.3cm]
Low $s$ exp. $\pi^+\pi^-$
	& $[2m_{\pi^\pm}-0.500]$   & $ 58.04\pm1.70\pm1.17$  
			& $ 56.03\pm1.60\pm0.28$ & $ +2.0\pm2.6$ \\
$\pi^+\pi^-      $  
	& $[0.500-1.800]$    & $450.16\pm4.89\pm1.57$  
			& $464.03\pm2.95\pm2.34$
                        & $-13.9\pm6.4$ \\
$\pi^0\gamma$, $\eta \gamma$\,$^(\footnotemark[1]{^)}$
	& $[0.500-1.800]$    & $  0.93\pm0.15\pm0.01$  & -  & - \\
$\omega$          
	& $[0.300-0.810]$    & $ 37.96\pm1.02\pm0.31$  & -  & - \\
$\pi^+\pi^-\pi^0$  {\footnotesize[below $\phi$]}
	& $[0.810-1.000]$    & $  4.20\pm0.40\pm0.05$  & -  & - \\
$\phi$  
	& $[1.000-1.055]$    & $ 35.71\pm0.84\pm0.20$  & -  & - \\
$\pi^+\pi^-\pi^0$  {\footnotesize[above $\phi$]}
	& $[1.055-1.800]$    & $  2.45\pm0.26\pm0.03$  & -  & - \\
$\pi^+\pi^-2\pi^0       $  
	& $[1.020-1.800]$    & $ 16.76\pm1.31\pm0.20$  
		& $ 21.45\pm1.33\pm0.60$
			& $ -4.7\pm1.8$ \\
$2\pi^+2\pi^-           $  
	& $[0.800-1.800]$    & $ 14.21\pm0.87\pm0.23$  
		& $ 12.35\pm0.96\pm0.40$
			& $ +1.9\pm2.0$ \\
$2\pi^+2\pi^-\pi^0        $  
	& $[1.019-1.800]$    & $  2.09\pm0.43\pm0.04$  & -  & - \\
$\pi^+\pi^-3\pi^0 $\,$^(\footnotemark[2]{^)}$  
	& $[1.019-1.800]$    & $  1.29\pm0.22\pm0.02$  & -  & - \\
$3\pi^+3\pi^-    $  
	& $[1.350-1.800]$    & $  0.10\pm0.10\pm0.00$  & -  & - \\
$2\pi^+2\pi^-2\pi^0       $  
	& $[1.350-1.800]$    & $  1.41\pm0.30\pm0.03$  & -  & - \\
$\pi^+\pi^-4\pi^0       $\,$^(\footnotemark[2]{^)}$    
	& $[1.350-1.800]$    & $  0.06\pm0.06\pm0.00$  & -  & - \\
$\eta${\footnotesize($ \rar\pi^+\pi^-\gamma$, $2\gamma$)}$\pi^+\pi^-$ 
	& $[1.075-1.800]$    & $  0.54\pm0.07\pm0.01$  & -  & - \\
$\omega${\footnotesize($\rar\pi^0\gamma$)}$\pi^{0}$
	& $[0.975-1.800]$    & $  0.63\pm0.10\pm0.01$  & -  & - \\
$\omega${\footnotesize($\rar\pi^0\gamma$)}$(\pi\pi)^{0}$
	& $[1.340-1.800]$    & $  0.08\pm0.01\pm0.00$  & -  & - \\
$K^+K^-            $  
	& $[1.055-1.800]$    & $  4.63\pm0.40\pm0.06$  & -  & - \\
$K^0_S K^0_L         $  
	& $[1.097-1.800]$    & $  0.94\pm0.10\pm0.01$  & -  & - \\
$K^0K^\pm\pi^\mp         $\,$^(\footnotemark[2]{^)}$    
	& $[1.340-1.800]$    & $  1.84\pm0.24\pm0.02$  & -  & - \\
$K\overline K\pi^0$\,$^(\footnotemark[2]{^)}$    
	& $[1.440-1.800]$    & $  0.60\pm0.20\pm0.01$  & -  & - \\
$K\overline K\pi\pi         $\,$^(\footnotemark[2]{^)}$    
	& $[1.441-1.800]$    & $  2.22\pm1.02\pm0.03$  & -  & - \\
$R=\sum{\rm excl.~modes}    $  
	& $[1.800-2.000]$    & $  8.20\pm0.66\pm0.10$  & -  & - \\
$R$ {\footnotesize[Data]}
	& $[2.000-3.700]$    & $ 26.70\pm1.70\pm0.03$  & -  & - \\
$J/\psi$         
	& $[3.088-3.106]$    & $  5.94\pm0.35\pm0.00$  & -  & - \\
$\psi(2S)$ 
	& $[3.658-3.714]$    & $  1.50\pm0.14\pm0.00$  & -  & - \\
$R$ {\footnotesize[Data]}  
	& $[3.700-5.000]$    & $  7.22\pm0.28\pm0.00$  & -  & - \\
$R_{udsc}$ {\footnotesize[QCD]}
	& $[5.000-9.300]$    & $  6.87\pm0.10\pm0.00$  & -  & - \\
$R_{udscb}$ {\footnotesize[QCD]}
	& $[9.300-12.00]$    & $  1.21\pm0.05\pm0.00$  & -  & - \\
$R_{udscbt}$  {\footnotesize[QCD]}
	& $[12.0-\infty]$    & $  1.80\pm0.01\pm0.00$  & -  & - \\[0.15cm]
\hline
&&&& \\[-0.3cm]
	&
			     & \mc{1}{r}{$696.3\pm6.2_{\rm exp}~~~$}  
			     & \mc{1}{l}{$711.0\pm5.0_{\rm exp}$} 
			     & \\
\rs{$\sum\;(e^+e^-\rightarrow\:$hadrons)}
 	& \rs{$[2m_{\pi^\pm}-\infty]$}
			     & \mc{1}{r}{$\pm\,3.6_{\rm rad\,}~~~$}  
			     & \mc{1}{r}{$\pm\,0.8_{\rm rad}\pm2.8_{\rm SU(2)}$} 
			     & \mc{1}{r}{\rs{$-14.7\pm7.9_{\rm tot}$}} 
	
\\[0.15cm]
 \hline
\end{tabular}
}
\end{center}
\vspace{-0.5cm}
{\footnotesize 
\begin{quote}
$^{1}\,$Not including $\omega$ and $\phi$ resonances (see Ref.~\cite{dehz}). 
\\ \noindent
$^{2}\,$Using isospin relations (see Ref.~\cite{dehz}). \\ \noindent
$^{3}\,$\ee\  data are used above 1.6~GeV (see Ref.~\cite{dehz}). \\ \noindent
\end{quote}
} 
\vspace{-0.75cm}
\caption{\label{tab_results}\em
	Summary of the \amuhadLO\ contributions from \ee\ 
        annihilation and $\tau$ decays. The uncertainties 
	on the vacuum polarization
	and FSR corrections are given as second errors in the individual 
        \ee\ contributions, while those from isospin breaking are 
        similarly given for the $\tau$ contributions. These 'theoretical'
        uncertainties are correlated among all channels, except in the
        case of isospin breaking which shows little correlation between 
        the $2\pi$ and $4\pi$ channels. The errors given 
        for the sums in the last line are from the experiment, the missing 
        radiative corrections in \ee\ and, in addition for $\tau$, SU(2)
        breaking.}
\end{table*}
The contributions from the different processes in their indicated 
energy ranges are listed in Table~\ref{tab_results}.
Wherever relevant, the two \ee- and $\tau$-based evaluations are given.
The discrepancies discussed above
are now expressed directly in terms of \amuhadLO, giving smaller
estimates for the \ee-based data set by 
$(-11.9\pm6.4_{\rm exp}\pm2.4_{\rm rad}\pm2.6_{\rm SU(2)}\,(\pm7.3_{\rm total}))~10^{-10}$ for the $\pi\pi$ channel and 
$(-2.8\pm2.6_{\rm exp}\pm0.3_{\rm rad}\pm1.0_{\rm SU(2)}\,(\pm2.9_{\rm total}))~10^{-10}$ for the sum of the $4\pi$ channels. 
The total discrepancy  
$(-14.7\pm6.9_{\rm exp}\pm2.7_{\rm rad}\pm2.8_{\rm SU(2)}\,(\pm7.9_{\rm total}))~10^{-10}$ amounts to 1.9 standard deviations. 
The difference could now be considered to be acceptable, however the systematic
difference between the \ee\ and $\tau$ $\pi\pi$ \sfs\ at high energies
precludes one from performing a straightforward combination of the 
two evaluations.

%
% -------------
%
\subsection{\it Results for $a_\mu$}
\label{sec_results_amu}

\begin{figure}[t]
\centerline{\psfig{file=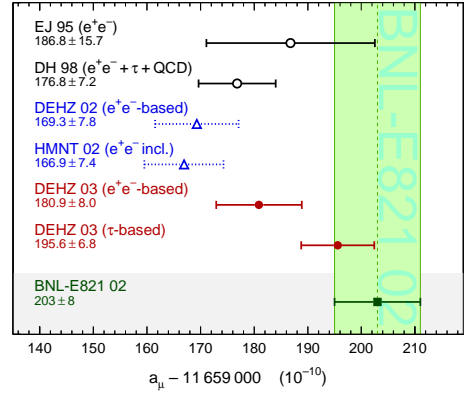,width=70mm}}
\caption[.]{\it Comparison of the results~(\ref{eq_smres}) with the 
	BNL measurement~\cite{bnl_2002}. Also given
	are our estimates~\cite{eidelman,dh98} obtained before 
	the CMD-2 data were available. For completeness, we show
	as triangles with dotted error bars the \ee-based 
	results~\cite{dehz,teubner} derived with the previously 
	published CMD-2 data~\cite{cmd2}.
	}
\label{fig:results}
\end{figure}

The results for the lowest order hadronic contribution are ($\times 10^{-10}$)
 \begin{eqnarray}
  a_{\mu, ee}^{\rm had,LO} &=&
       696.3\pm6.2_{\rm exp}\pm3.6_{\rm rad} \\
  a_{\mu, \tau}^{\rm had,LO} &=&
       711.0\pm5.0_{\rm exp}\pm0.8_{\rm rad}
			\pm2.8_{\rm SU(2)} 
 \end{eqnarray}
Adding to these the QED, higher-order hadronic, light-by-light scattering and
weak contributions as given in Section~\ref{anomaly},
we obtain the SM predictions ($\times 10^{-10}$, with an additional common
uncertainty of $\pm3.5_{\rm LBL}\pm0.4_{\rm QED+EW}$)
\beqns
\label{eq_smres}
  a_{\mu, ee}^{\rm SM} &=& 11\,659\,180.9\pm7.2_{\rm had,LO} \\
  a_{\mu, \tau}^{\rm SM} &=& 	11\,659\,195.6\pm5.8_{\rm had,LO}
\eeqns
These values can be compared to the present measurement~(\ref{eq:bnlexp}). 
Adding experimental and theoretical errors 
in quadrature, the differences between measured and computed values 
are found to be ($\times 10^{-10}$, with an additional common
uncertainty of $\pm3.5_{\rm other}\pm8.0_{\rm exp}$ from contributions 
other than hadronic vacuum polarization and the BNL g-2 experimental error)
\beq
\label{eq:diffbnltheo}
 \begin{array}{rcll}
  a_\mu^{\rm exp}-a_{\mu, ee}^{\rm SM} &=&
	22.1\pm7.2_{\rm had,LO} \\
  a_\mu^{\rm exp}-a_{\mu, \tau}^{\rm SM} &=&
	 7.4\pm5.8_{\rm had,LO}
 \end{array}
\eeq
where the error quoted is specific to each approach.
The differences~(\ref{eq:diffbnltheo})
correspond to 1.9 and 0.7 standard deviations, respectively.
A graphical comparison of the results~(\ref{eq_smres}) with the 
experimental value is given in Fig.~\ref{fig:results}. 
Also shown are our estimates~\cite{eidelman,dh98},
obtained before the CMD-2 and the new $\tau$ data were available 
(see discussion below), and the \ee-based
evaluations of Refs.~\cite{dehz,teubner},
obtained with the previously published, uncorrected CMD-2 data~\cite{cmd2}.

%\vfill
%\pagebreak

%
% ----------------------------- Discussion ------------------------------
%
\section{Discussion}
\label{sec_discuss}

\begin{figure}[p]
\centerline{\psfig{file=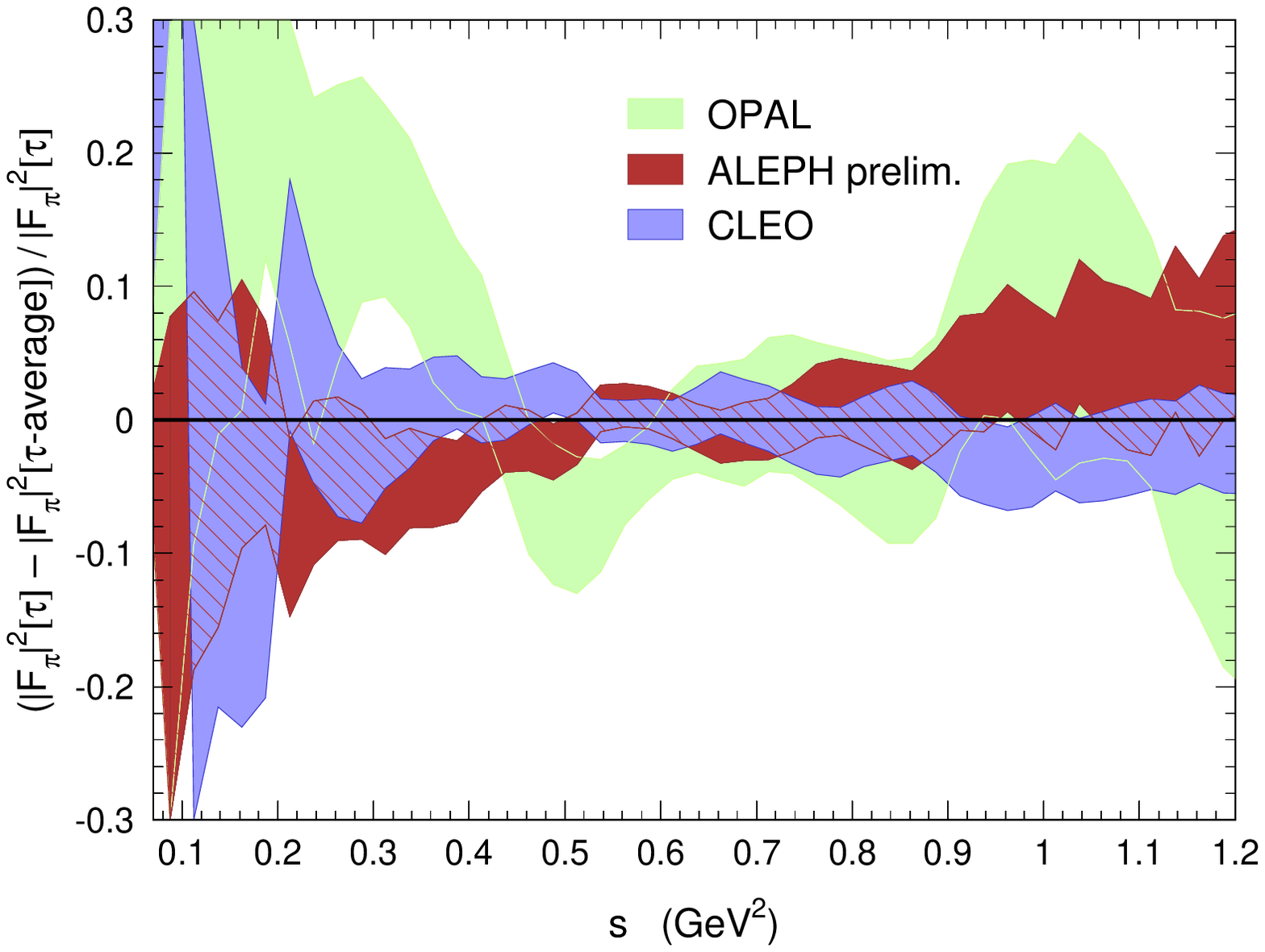,width=70mm}}
\caption[.]{\it Relative comparison of the $\pi^+\pi^-$ \sfs\
    	extracted from $\tau$ data from different experiments, 
	expressed as a ratio to the average $\tau$ \sf.}
\label{comp_tau}
%\end{figure}
%\begin{figure}[t]
\vspace{0.6cm}
\centerline{\psfig{file=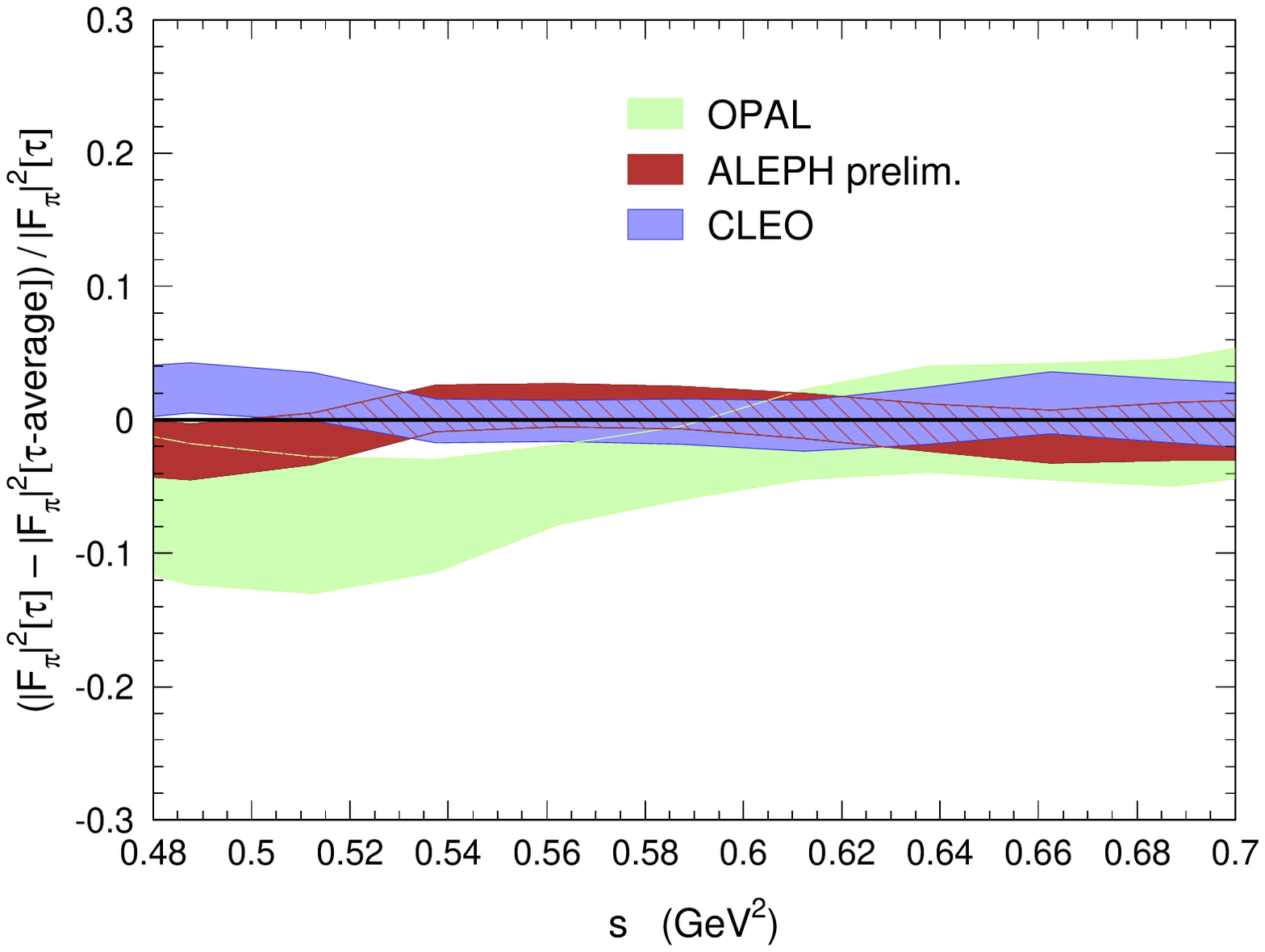,width=70mm}}
\caption[.]{\it Relative comparison in the $\rho$ region of the 
	$\pi^+\pi^-$ \sfs\ extracted from $\tau$ data, 
	expressed as a ratio to the average $\tau$ \sf.}
\label{comp_tau_zoom}
\end{figure}

Although the new corrected CMD-2 $\pi^+\pi^-$ results are now consistent with 
$\tau$ data for the energy region below 850~MeV, the remaining discrepancy
for larger energies is unexplained at present. Hence, one could 
question the validity of either \ee\ data with their large radiative 
corrections, $\tau$ data, or the isospin-breaking corrections 
applied to $\tau$ data. We shall briefly discuss these points below.
\bei
\item The CMD-2 experiment is still the only one claiming systematic 
accuracies well below $1\%$. It is thus difficult to confront their
data with results from other experiments. Whereas the measurements 
from OLYA are systematically lower than the new CMD-2
results in the peak region, there is a trend of agreement above, 
as seen in Fig.~\ref{fig_2pi_comp}. This behaviour appears to be confirmed
by preliminary data from the KLOE experiment at Frascati using the radiative
return method from the $\phi$ resonance~\cite{kloe_prel}. 
\item The most precise results on the $\tau$ $\pi\pi$ spectral function come
from the ALEPH and CLEO experiments, operating in completely different
physical environments. On the one hand, the main uncertainty in CLEO 
originates from the knowledge of the relatively low selection efficiency, 
a consequence of the large non-$\tau$ hadronic background, while the  
mass spectrum is measured with little distortion and good resolution.
On the other hand, ALEPH has both large efficiency and small background, 
the main uncertainty coming from the $\pi^0$ reconstruction close to 
the charged pion, necessitating to unfold the measured spectrum
from detector resolution and acceptance effects. A comparison of the 
$\tau$ spectral functions from ALEPH, CLEO and OPAL is given 
in Fig.~\ref{comp_tau}, and in Fig.~\ref{comp_tau_zoom} 
for the $\rho$ peak region. Agreement is observed within quoted errors, 
in particular in the high mass region,
although CLEO results are a bit closer to \ee\ data there. Overall,
the $\tau$ data appear to be consistent.

\item The last point concerns isospin corrections applied to the $\tau$
spectral functions.
The basic components entering SU(2) breaking are well identified. 
The long-distance radiative corrections and the quantitative effect of loops 
have been addressed by the analysis of Ref.~\cite{ecker2} showing that 
the effects are small. The overall effect of the isospin-breaking 
corrections (including FSR) applied to the $\tau$ $\pi\pi$ data, 
expressed in relative terms, is $(-1.8\pm0.5)\%$. Its largest contribution 
($-2.3\%$) stems from the uncontroversial short-distance electroweak 
correction~\cite{marciano-sirlin}.
One could question the validity of the chiral model used. The authors of
Ref.~\cite{ecker2} argue that the corrections are insensitive to the 
details of their model and essentially depend only on the shape of the
pion form factor. As the latter is known from experiment to adequate
accuracy, it seems difficult to find room for a $\sim 10\%$ effect as
observed experimentally. Nevertheless, considering the situation
regarding the first two experimental points, it would seem worthwhile 
to invest more theoretical work into the problem of isospin breaking.

The particular point of the $\rho^- - \rho^0$ mass splitting deserves some
further discussion. This possibility was considered from the beginning when
it was proposed to use $\tau$ spectral functions to compute vacuum
polarization integrals~\cite{adh}. However, at that time, the experimental
investigation led to a splitting consistent with 0 within 1.1~MeV, supported 
by theoretical investigations~\cite{bijnens} which indicated a value less 
than 0.7~MeV. Consequently,in further analyses,
we assumed $m_{\rho^0}=m_{\rho^-}$ to hold within
an uncertainty of 1~MeV. A re-analysis of this question with the current 
more precise data on \ee\ and $\tau$ spectral functions~\cite{md_tausf} leads 
to the conclusion that the mass splitting favoured by the data is 
$m_{\rho^-}-m_{\rho^0}=(2.3\pm0.8$~MeV. A similar conclusion was more recently
reached~\cite{gj}. Unfortunately, it is not yet clear whether such a result
must be taken as definitive evidence for a mass splitting and if the 
corresponding correction must be applied to the $\tau$ spectral function. 
The significance of the result is still limited, but the most worrisome 
problem is the fact that the \ee\ and the mass-corrected $\tau$ spectral 
functions still disagree in magnitude by $3.3\%$~\cite{md_tausf}. Therefore, 
while  the consideration of the mass splitting improves the line shape 
comparison, it leaves us with a major normalization discrepancy. In other 
words, the $\rho^-$ mass correction spreads the large local difference 
observed in the 0.85-1.~GeV range almost uniformly across the mass spectrum. 
We thus disagree with the conclusions reached in Ref.~\cite{gj}
that taking into account the $\rho$ mass difference solves the current 
discrepancy. If we take the point of view that the remaining difference is 
basically a normalization problem (either from the data or the isospin breaking
corrections), then correcting for the apparent mass difference increases
the $\tau$-based estimate of $a_\mu^{\rm had,LO}$ by $5.4~10^{-10}$, bringing
it even closer to the BNL result, but further away from the \ee-based
estimate for a total of $2.5~\sigma$. More experimental investigation with
CMD-2, KLOE and \babar\ is needed to consolidate the \ee\ picture.
\eei

%
% --------------------------- Conclusions -----------------------------
%
\section{Conclusions}

An update of our analysis of the lowest-order hadronic vacuum 
polarization contribution to the muon anomalous magnetic moment 
has been performed following a reevaluation by the CMD-2 Collaboration
of their \ee\ annihilation cross sections. Part of the previous 
discrepancy between the \ee\ and $\tau$ $\pi\pi$ spectral functions 
has now disappeared so that the corresponding evaluations of the 
lowest-order hadronic polarization contribution to the muon magnetic 
anomaly are closer. However, incompatible measurements remain
between 0.8 and 1~GeV so that we do not proceed with an average 
of the two evaluations.
The \ee- and $\tau$-based predictions are respectively 1.9 and 0.7 standard
deviations below the direct measurement from the g-2 Collaboration at BNL.
Considering the $\pi\pi$ discrepancy from the point of view of the 
$\rho^- - \rho^0$ mass splitting turns out to increase the difference 
between the two estimates.
The forthcoming results from radiative return with KLOE~\cite{kloe_prel}
and \babar~\cite{md_babar} will be important to sort out possibly remaining 
problems in the $\pi\pi$ and $4\pi$ \sfs.

\section*{Acknowledgments}

I would like to thank Simon Eidelman, Andreas H\"ocker,
and Zhiqing Zhang for the fruitful collaboration, and Gerhard Ecker
for stimulating discussions. 
Congratulations to Marco Incagli and
Graziano Venanzoni for organizing a very stimulating workshop.


\begin{thebibliography}{999}

\bibitem{eidelman}    S.~Eidelman and F.~Jegerlehner,
                      {\it Z. Phys.} {\bf C67} (1995) 585.
\bibitem{adh}         R.~Alemany, M.~Davier and A.~H\"ocker,
                      {\it Eur.Phys.J.} {\bf C2} (1998) 123.
\bibitem{aleph_vsf}   R.~Barate {\it et al.}, (ALEPH Collaboration),
                      {\it Z. Phys.} {\bf C76} (1997) 15.
\bibitem{aleph_asf}   R.~Barate {\it et al.}, (ALEPH Collaboration),
                      {\it Eur. J. Phys.} {\bf C4} (1998) 409.
\bibitem{dh97}        M.~Davier and A.~H\"ocker,
                      {\it Phys. Lett.} {\bf B419} (1998) 419.
\bibitem{steinhauser} J.H.~K\"uhn and M.~Steinhauser,
                      {\it Phys. Lett.} {\bf B437} (1998) 425.
\bibitem{martin}      A.D.~Martin and D.~Zeppenfeld,
                      {\it  Phys. Lett.} {\bf B345} (1995) 558.
\bibitem{groote}      S.~Groote {\it et al.},
                      {\it Phys. Lett.} {\bf B440} (1998) 375.
\bibitem{dh98}        M.~Davier and A.~H\"ocker,
                      {\it Phys. Lett.} {\bf B435} (1998) 427.
\bibitem{dehz}        M.~Davier, S.~Eidelman, A.~H\"ocker and Z.~Zhang,
                      {\it Eur.Phys.J.} {\bf C27} (2003) 497.
\bibitem{cmd2}        R.R. Akhmetshin {\it et al.} (CMD-2 Collaboration),
                      {\it Phys.Lett.} {\bf B527} (2002) 161.
\bibitem{aleph_new}   M.~Davier and C.~Yuan,
                      {\it Nucl.Phys.} B (Proc.Suppl.) {\bf 123} (2003) 47.
\bibitem{ecker1}      V.~Cirigliano, G.~Ecker and H.~Neufeld,
                      {\it Phys. Lett.} {\bf B513} (2001) 361.
\bibitem{ecker2}      V.~Cirigliano, G.~Ecker and H.~Neufeld, 
	              {\em JHEP} {\bf 0208} (2002) 002.
\bibitem{teubner}     K.~Hagiwara, A.D.~Martin, D.~Nomura and
		      T.~Teubner, {\em Phys. Lett.} {\bf B557} (2003) 69.
\bibitem{bnl_2002}    G.W.~Bennett {\it et al.} (Muon g-2 Collaboration),
                      {\em Phys. Rev. Lett.} {\bf 89} (2002) 101804; 
		      Erratum-ibid. {\bf 89} (2002) 129903.
\bibitem{dehz03}      M.~Davier, S.~Eidelman, A.~H\"ocker and Z.~Zhang,
                      hep-ph/0308213, to appear in {\it Eur.Phys.J.}
\bibitem{cmd2_new}    R.~Akhmetshin {\it et al.} (CMD-2 Collaboration),
                      hep-ex/0308008 (2003).
\bibitem{L3_hpi0}     P.~Achard {\it et al.} (L3 Collaboration),
                      CERN-EP/2003-019 (May 2003).
\bibitem{hughes}      V.W.~Hughes and T.~Kinoshita,
                      {\it Rev. Mod. Phys.} {\bf 71} (1999) 133.
\bibitem{cm}          A.~Czarnecki and W.J.~Marciano,
                      {\it Nucl. Phys. (Proc. Sup.)} {\bf B76} (1999) 245.
\bibitem{kino_nio}    T.~Kinoshita and M.~Nio,
                      {\it Phys. Rev. Lett.} {\bf 90} (2003) 02803.
\bibitem{nyff}        A.~Nyffeler, hep-ph/0305135.
\bibitem{krause2}     B.~Krause, {\it Phys. Lett.} {\bf B390} (1997) 392.
\bibitem{amuweak}     A.~Czarnecki, W.J.~Marciano and A.~Vainshtein,
		      {\em Phys. Rev.} {\bf D67} (2003) 073006;
                      A.~Czarnecki, B.~Krause and W.J.~Marciano,
                      {\it Phys. Rev. Lett.} {\bf 76} (1995) 3267;
                      {\it Phys. Rev.} {\bf D52} (1995) 2619;
		      R.~Jackiw and S.~Weinberg, 
                      {\it Phys. Rev.} {\bf D5} (1972) 2473;
                      S.~Peris, M.~Perrottet and E.~de Rafael,
                      {\it Phys. Lett.} {\bf B355} (1995) 523;
                      M.~Knecht {\it et al.}, 
		      {\em JHEP} {\bf 0211} (2002) 003.
\bibitem{knecht_light}M.~Knecht {\it et al.},
                      {\it Phys.Rev.} {\bf D65} (2002) 073034. 
\bibitem{kino_light_cor}   
		      M.~Hayakawa and T.~Kinoshita, 
		      Erratum {\em Phys. Rev.} {\bf D66} (2002) 019902;
		      ibid. {\bf D57} (1998) 465.
\bibitem{bij_light_cor}    
	              J.~Bijnens, E.~Pallante and J.~Prades,
                      {\it Nucl.Phys.} {\bf B626} (2002) 410.
\bibitem{rafael}      M.~Gourdin and E.~de~Rafael, 
                      {\it Nucl. Phys.} {\bf B10} (1969) 667;
                      S.J.~Brodsky and E.~de Rafael, 
                      {\it Phys. Rev.} {\bf 168} (1968) 1620.
\bibitem{snd_omega}   M.~N.~Achasov {\it et al.}, (SND Collaboration),
                      hep-ex/0305049.
\bibitem{snd_4pi}     M.~N.~Achasov {\it et al.}, (SND Collaboration),
                      {\em J. of Exp. and Theor. Physics,}
                      {\bf 96} (2003) 789.
\bibitem{cleo_2pi}    S.~Anderson {\it et al.} (CLEO Collaboration),
                      {\it Phys.Rev.} {\bf D61} (2000) 112002.
\bibitem{opal_2pi}    K. Ackerstaff {\it et al.} (OPAL Collaboration),
                      {\it Eur. Phys. J.} {\bf C7} (1999) 571.
\bibitem{cleo_bpipi0} M.~Artuso {\it et al.} (CLEO Collaboration),
                      {\it Phys. Rev. Lett.} {\bf 72} (1994) 3762.
\bibitem{opal_bpipi0} K.~Ackerstaff {\it et al.} (OPAL Collaboration),
                      {\it Eur. Phys. J.} {\bf C4} (1998) 93.
\bibitem{aleph_ksum}  R.~Barate {\em et al.} (ALEPH Collaboration),
                      {\it Eur. Phys. J.} {\bf C11} (1999) 599.
\bibitem{cleo_kpi0}   M.~Battle {\it et al.} (CLEO Collaboration),
                      {\it Phys. Rev. Lett.} {\bf 73} (1994) 1079. 
\bibitem{kloe_prel}   B.~Valeriani, {\it Preliminary results from KLOE on the
                      radiative return}, these proceedings.
\bibitem{marciano-sirlin}  
		      W.~Marciano and A.~Sirlin, {\it Phys. Rev. Lett.}
                      {\bf 61} (1988) 1815.
\bibitem{bijnens}     J.~Bijnens and P.~Gosdzinsky, 
                      {\it Phys. Lett.} {\bf B388} (1996) 203.
\bibitem{md_tausf}    M.~Davier, {\it Spectral functions from hadronic 
                      $\tau$ decays}, these proceedings.
\bibitem{gj}          S.~Ghozzi and F.J.~Jegerlehner, hep-ph/0310181
\bibitem{md_babar}    M.~Davier, {\it R measurement through ISR with BaBar}, 
                      these proceedings.
\end{thebibliography}
\end{document}